\documentclass[cameraready]{Interspeech}

\title{I'll Keep an Ear Out: Teaching AudioLLMs Proactive Audio Assistance}

\author[affiliation={1}, equalcontribution]{Amit Kumar Singh}{Yadav}
\author[affiliation={1}, equalcontribution]{Ritvik}{Shrivastava}
\author[affiliation={1}]{Xuan}{Zhang}
\author[affiliation={1}]{Seungwhan}{Moon}
\author[affiliation={1}]{Shashank}{Jain}
\author[affiliation={1}]{Pinar}{Donmez}
\author[affiliation={1}]{Babak}{Damavandi}

\address{
    $^1$ Meta Reality Labs, USA
}

\email{
    \{amit99, ritvik, xuanzhang216, shanemoon, shajain, pinared, babakd\}@meta.com
}


\usepackage{multirow}
\usepackage{makecell}
\usepackage{subcaption}

\begin{document}

\maketitle

\begin{abstract}

Audio large language models (AudioLLMs) operate reactively, responding only when queried. We introduce proactive audio assistance, where an AudioLLM monitors an audio stream and autonomously decides when to alert the user from a single natural-language intent, motivated by wearable applications for Deaf and Hard of Hearing users. We propose Interrupt and Silent Modeling (ISM), a model-agnostic paradigm that embeds proactive decisions into LLM decoding via two special tokens: \texttt{<interrupt>} and \texttt{<silent>}, capturing four states: onset detection, sustained-relevance triggering, irrelevance suppression, and de-duplication. Applied to Qwen2-Audio-7B, ISM achieves 99.6\% interrupt F1 and perfect de-duplication recall on ESC-50. On noisy Epic-Sounds kitchen audio, ISM achieves the highest interrupt F1 without domain-specific training, the only method maintaining strong onset detection without over-triggering or over-suppression. Streaming evaluation confirms real-time viability with 3.5-second average latency.


\end{abstract}

\section{Introduction}

Audio large language models (AudioLLMs) have advanced rapidly in understanding and reasoning about acoustic signals~\cite{chu2024qwen2audio, ghosh2024gama, he2024meralion}. These systems classify sounds, answer questions about audio, and generate natural language descriptions. However, they share a fundamental limitation: they are \emph{purely reactive}. Each interaction requires the user to formulate a query, wait for a response, and repeat for every new event. This paradigm is inadequate for scenarios requiring continuous monitoring, such as alerting a Deaf or Hard of Hearing (DHH) individual when a specific sound occurs~\cite{matthews2006dhh, glasser2017dhh}.

We envision proactive AudioLLMs that determine both \emph{when} and \emph{how} to assist without requiring a query before every response. We identify three categories of proactive triggers: (i)~\emph{explicit} requests (e.g., ``watch for someone knocking on my door''), (ii)~\emph{implicit} cues (e.g., detecting hesitation or confusion in user speech), and (iii)~\emph{semantic-driven} assistance (e.g., recognizing that a traveler at an airport gate could benefit from a boarding announcement alert). In this work, we focus on the explicit watch-out intent scenario as the foundational case.

\begin{figure}[t]
  \centering
  \includegraphics[width=\linewidth]{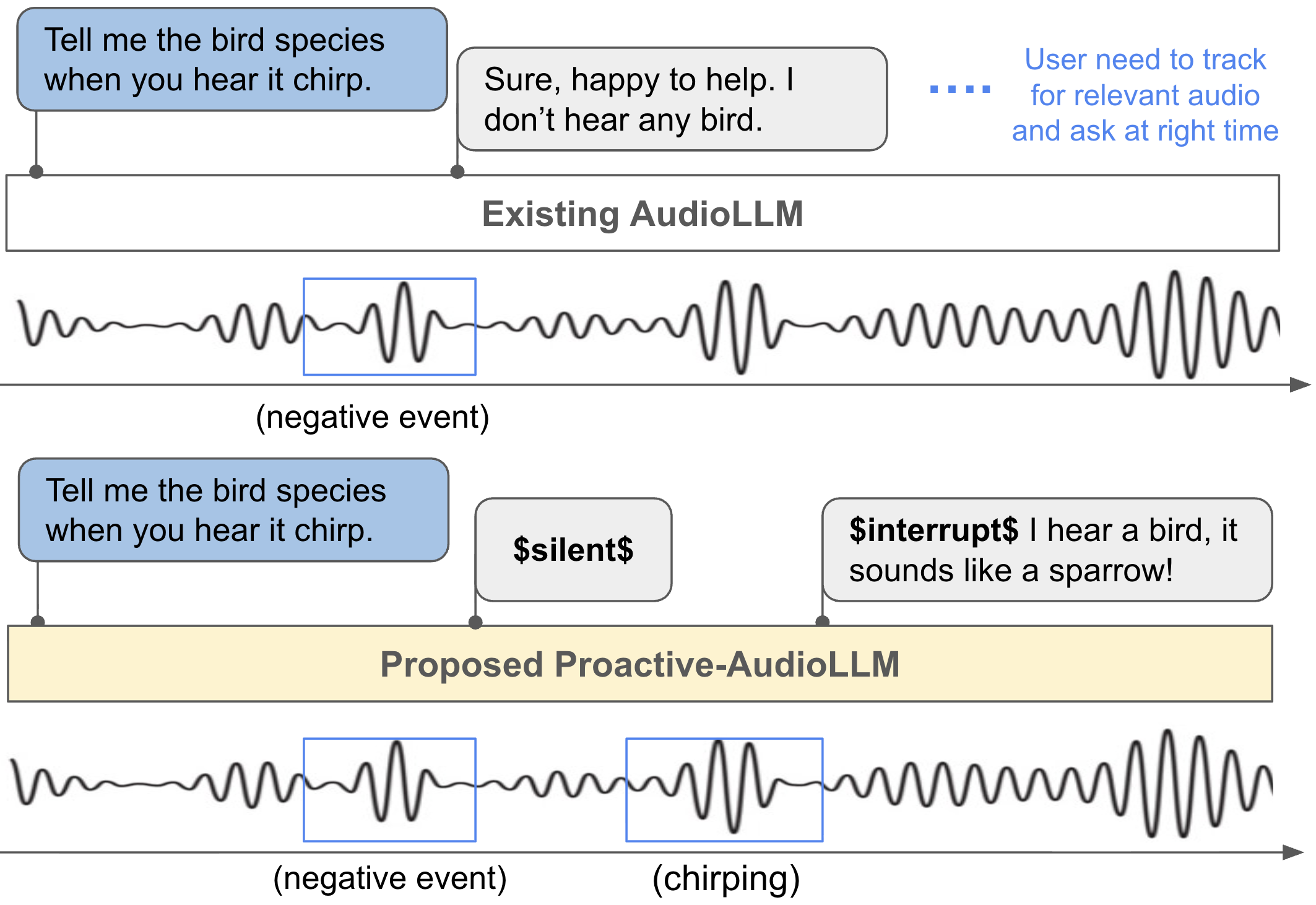}
  \caption{Reactive vs.\ proactive audio assistance. Existing AudioLLMs (top) require a query for every event. Our proactive approach (bottom) takes a single watch-out intent and autonomously decides when to intervene.}
  \label{fig:reactive_vs_proactive}
\end{figure}

We introduce \emph{proactive audio assistance} (Figure~\ref{fig:reactive_vs_proactive}), in which the user specifies a single watch-out intent and the AudioLLM autonomously monitors an incoming audio stream, deciding at each moment whether to \emph{interrupt} with a notification or remain \emph{silent}. This addresses a critical gap in assistive technology: conventional sound classification systems for DHH users~\cite{matthews2006dhh, glasser2017dhh} monitor a fixed set of sound classes without modeling user intent or interaction history, leading to notification fatigue from repeated alerts for the same ongoing event. Our approach treats de-duplication as a first-class modeling objective - the system maintains awareness of its notification history and suppresses redundant alerts for sustained sound events.

Proactive audio assistance poses significant challenges beyond standard classification. The model must perform temporal reasoning over streaming audio to detect the onset of relevant events, distinguish first occurrences from continuations of already-reported events, handle noisy and overlapping acoustic scenes without false alarms, and make these decisions causally, using only past and present context, under real-time latency constraints.

%
%

\noindent \textbf{Our contributions} are as follows:
\begin{itemize}
    \item We formalize proactive audio assistance as a new task for AudioLLMs, defining four decision states: onset interruption, sustained-relevance interruption, irrelevance silence, and de-duplication silence.
    \item We propose Interrupt and Silent Modeling (ISM), embedding proactive decisions directly into LLM decoding via two special tokens added to the vocabulary.
    \item We define proactive evaluation metrics and a streaming protocol for real-time assessment.
    \item We demonstrate near-perfect proactive performance on ESC-50~\cite{piczak2015esc} (99.6\% interrupt F1, 100\% de-duplication recall) and robust zero-shot interrupt transfer to Epic-Sounds~\cite{huh2023epic} using Qwen2-Audio-7B~\cite{chu2024qwen2audio} - showing that proactive behavior is achievable through ISM alone, without architectural changes, and that ISM is model-agnostic by design.
\end{itemize}

\section{Related work}

\textbf{Audio understanding.}
Environmental sound classification has progressed from handcrafted spectral features to deep learning. The Audio Spectrogram Transformer (AST)~\cite{gong2021ast} achieves strong accuracy on ESC-50~\cite{piczak2015esc} via large-scale pretraining on AudioSet~\cite{gemmeke2017audioset}. Self-supervised methods such as BEATs~\cite{chen2023beats} and HTS-AT~\cite{chen2022htsat} further push performance, with BEATs achieving 98.1\% on ESC-50. Epic-Sounds~\cite{huh2023epic} extends evaluation to egocentric kitchen audio with 44 temporally localized event classes recorded in noisy, real-world conditions.

\noindent \textbf{AudioLLMs.}
Recent AudioLLMs integrate audio encoders with LLMs for open-ended audio understanding~\cite{chu2024qwen2audio, ghosh2024gama, he2024meralion}. Qwen2-Audio~\cite{chu2024qwen2audio} pairs Whisper-large-v3~\cite{radford2023whisper} with a 7B language model, supporting both speech and non-speech audio tasks. GAMA~\cite{ghosh2024gama} extends audio reasoning with complex chain-of-thought capabilities. These systems excel at query-driven tasks but lack mechanisms for autonomous, intent-aware monitoring - each response requires an explicit user prompt.

\noindent \textbf{Sound awareness for DHH users.}
Wearable sound awareness systems for DHH individuals have been explored extensively~\cite{matthews2006dhh, glasser2017dhh}. Conventional approaches continuously monitor a fixed set of sound classes and generate alerts for every detected event. This leads to notification fatigue, as users cannot express intent (which sounds matter) or suppress repeated alerts for ongoing events. Our approach addresses both limitations through user-specified intents and history-aware de-duplication.

\noindent \textbf{Proactive and anticipatory AI.}
Proactive behavior has been explored in conversational agents~\cite{deng2023proactive} and video understanding ~\cite{kundu2026planwatchrecoverbenchmark}. VideoLLM-Online~\cite{chen2024videollmonline} proactively narrates key activities in video streams, and Mirai~\cite{lee2024mirai} uses image sequences for timely behavioral nudges.  No prior work addresses proactive assistance for AudioLLMs. We formalize proactive audio monitoring as a new task for AudioLLMs, define a four-state decision framework, and introduce ISM as a model-agnostic training paradigm to enable it.

\section{Method}

\subsection{Task formulation}

Given a user-specified watch-out intent $q$ (e.g., ``alert me if a dog barks'') and a continuous audio stream, the model decides at each time step whether to interrupt or remain silent. We define four decision states based on audio content and interaction history:

\noindent \textbf{Interrupt Type-1 (I1):} The audio transitions from irrelevant to relevant content relative to $q$, modeling \emph{onset detection} - the first moment a relevant sound appears after irrelevant audio.

\noindent \textbf{Interrupt Type-2 (I2):} The entire observed audio window contains relevant content with no preceding irrelevant context, modeling \emph{sustained relevance}. This addresses cases where the model missed the initial onset and must still trigger. Empirically, adding I2 training improves I2 recall by ${\sim}$10 percentage points (Section~\ref{sec:ablation}).

\noindent \textbf{Silent Type-1 (S1):} The audio is unrelated to $q$; the model remains silent.

\noindent \textbf{Silent Type-2 (S2):} The audio is relevant to $q$, but the model has already notified the user (recorded in conversation history). The model remains silent to avoid duplicate alerts (\emph{de-duplication}).

\subsection{Architecture}

\begin{figure*}[t]
    \centering
    \begin{subfigure}[t]{0.5\textwidth}
      \centering
      \includegraphics[width=\linewidth,height=5cm,keepaspectratio]{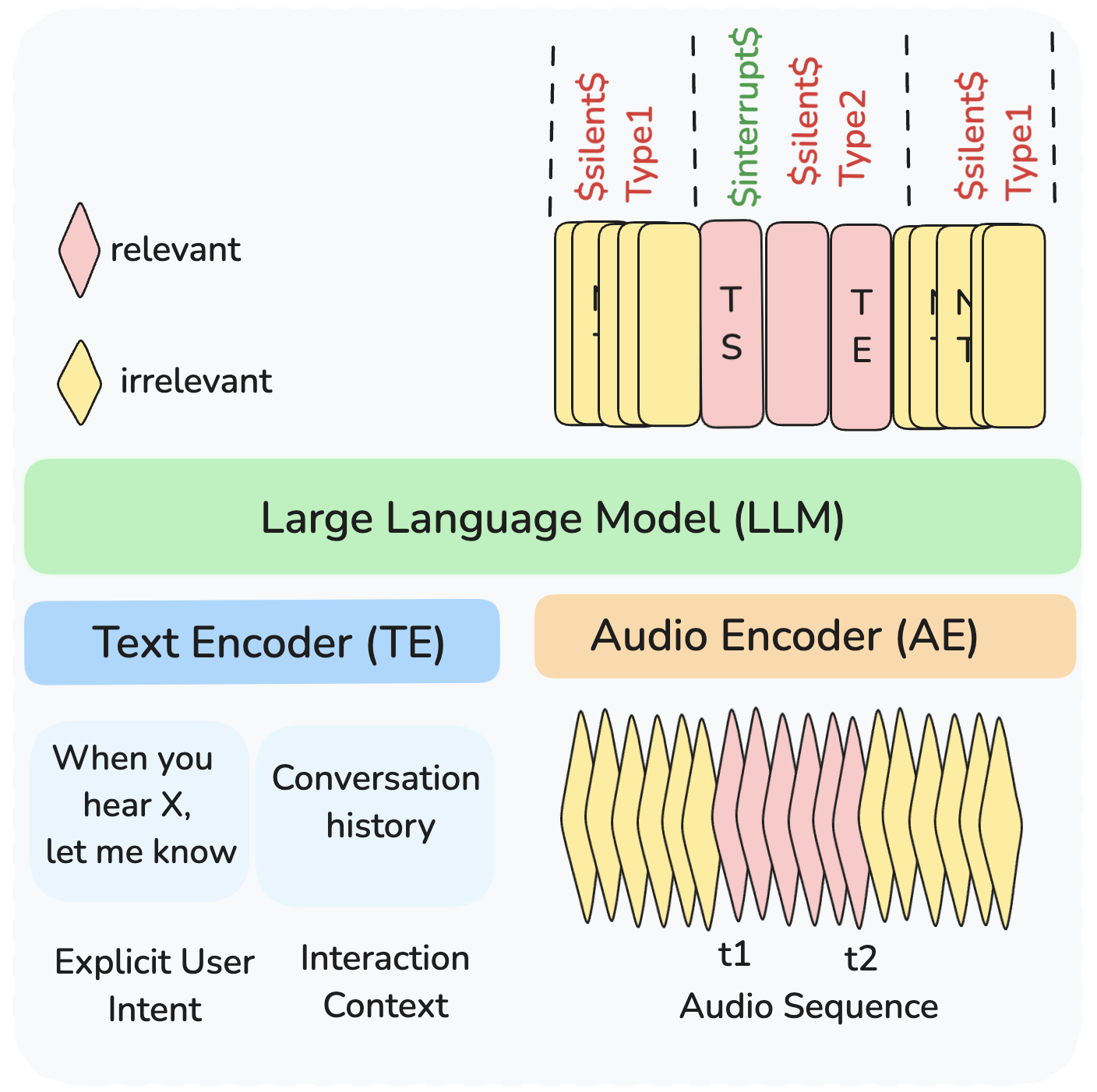}
      \caption{PALLM architecture.}
      \label{fig:pallm_arch}
    \end{subfigure}\hfill
    \begin{subfigure}[t]{0.48\textwidth}
      \centering
      \includegraphics[width=\linewidth,height=5cm,keepaspectratio]{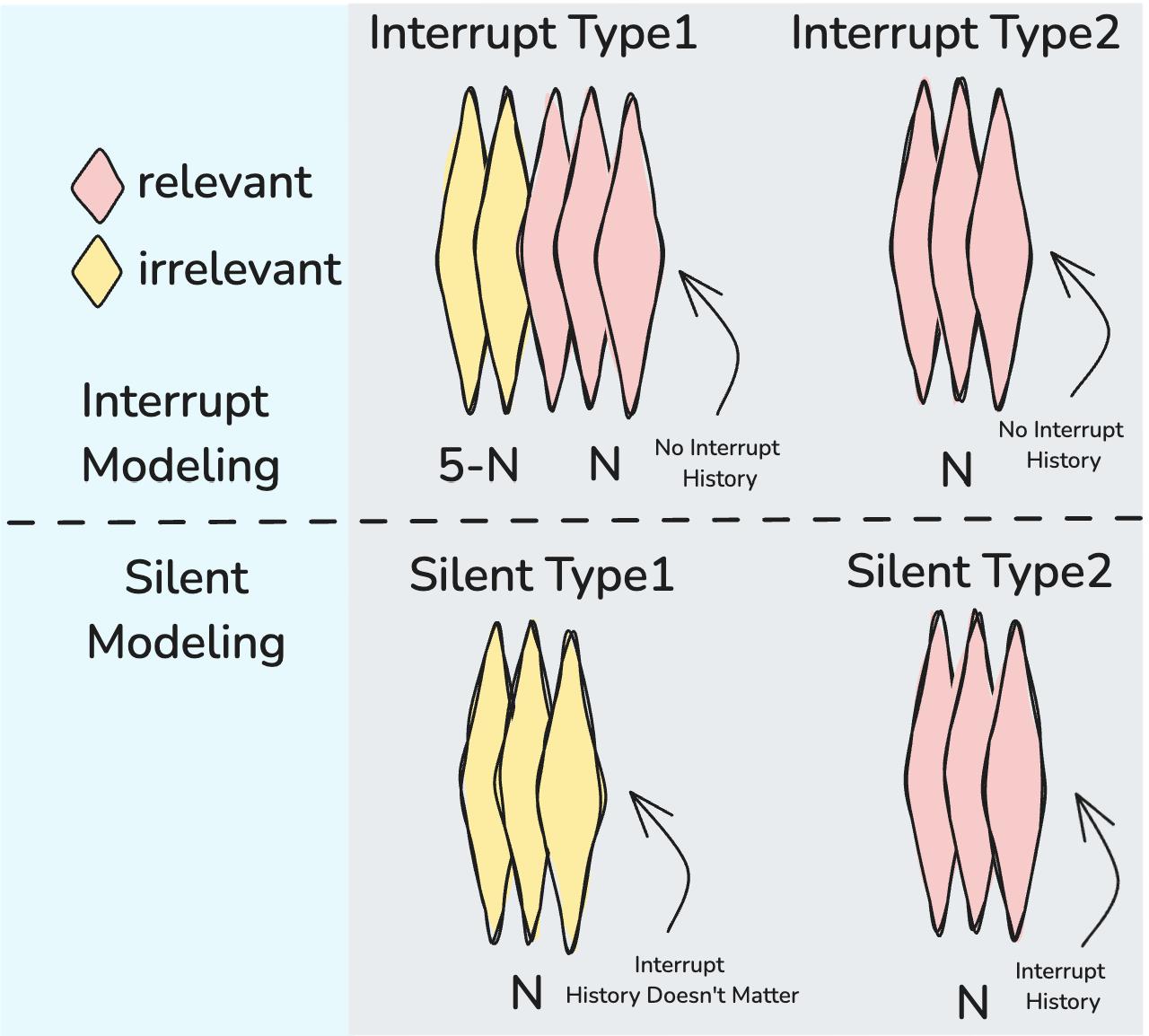}
      \caption{Interrupt and Silent Modeling (ISM).}
      \label{fig:ism}
    \end{subfigure}
    \caption{(a) PALLM takes a single watch-out intent and continuously encodes audio, decoding either \texttt{<interrupt>} or \texttt{<silent>}. (b) ISM training with four decision states. Conversation history enables de-duplication (S2).}
    \label{fig:architecture}
\end{figure*}

Our Proactive Audio Large Language Model (PALLM), shown in Figure~\ref{fig:pallm_arch}, is built on Qwen2-Audio-7B~\cite{chu2024qwen2audio}, comprising a Whisper-large-v3 audio encoder~\cite{radford2023whisper} and a 7B-parameter language model (${\sim}$8B total parameters). Audio is resampled to 16~kHz and converted to 128-dimensional mel-spectrograms (25~ms window, 10~ms hop). A pooling layer with stride~2 yields ${\sim}$40~ms temporal resolution per token. The user's watch-out intent and interaction history are tokenized using Qwen2-Audio's text tokenizer - no separate text encoder module is required.

We deliberately use a standard AudioLLM backbone to demonstrate that proactive behavior is enabled by ISM, not by architectural novelty. ISM is model-agnostic: it requires only vocabulary extension and autoregressive decoding - properties shared by all current AudioLLMs, including GAMA~\cite{ghosh2024gama} and MERaLiON~\cite{he2024meralion}. No changes to the ISM paradigm itself are needed to apply it to a different backbone; the same two-token vocabulary extension, training data construction, and loss formulation transfer directly.

%
%

\subsection{Interrupt and Silent Modeling (ISM)}

ISM extends the LLM's vocabulary with two special tokens: \texttt{<interrupt>} and \texttt{<silent>}. These enable proactive decisions within the standard autoregressive decoding framework (Figure~\ref{fig:ism}). Training proceeds in two stages: (1)~reactive supervised fine-tuning (SFT) for sound classification, and (2)~proactive SFT for interrupt and silent modeling.

\noindent \textbf{Reactive SFT.}
We first fine-tune the model on sound classification using sequence-to-sequence learning with cross-entropy loss. Given audio input and a classification query, the model learns to generate the correct class label. This establishes the acoustic understanding required for proactive decisions.

\noindent \textbf{Proactive SFT - training data construction.}
For each audio sample with known event boundaries $[t_1, t_2]$, we construct training instances for all four decision states. I1 instances pair irrelevant audio preceding $t_1$ with relevant audio after $t_1$, simulating an onset event during streaming. I2 instances contain only relevant audio within $[t_1, t_2]$, with no preceding irrelevant context. S1 instances sample audio fully outside $[t_1, t_2]$; multiple S1 instances per sample train the model to consistently remain silent for irrelevant input. S2 instances contain relevant audio but include a conversation history entry indicating a prior notification, training the model to suppress duplicate alerts.

\noindent \textbf{Proactive SFT - loss formulation.}
For interrupt instances, the target sequence is [\texttt{<interrupt>}, response tokens]; for silent instances, [\texttt{<silent>}]. The losses are:
\begin{align}
    \mathcal{L}_I &= -\sum_{k=1}^{N} \log \mathrm{softmax}(o_k) \cdot i_k \label{eq:interrupt} \\
    \mathcal{L}_S &= -\log \mathrm{softmax}(o_1) \cdot s_1 \label{eq:silent} \\
    \mathcal{L}_P &= \sum_{j=1}^{N_I} \mathcal{L}_I^{(j)} + \sum_{j=1}^{N_S} \mathcal{L}_S^{(j)} \label{eq:proactive}
\end{align}
where $o_k$ is the logit vector at position $k$, $i_k$ and $s_1$ are the corresponding one-hot target vectors, $N$ is the target sequence length, and $\mathcal{L}_I^{(j)}$/$\mathcal{L}_S^{(j)}$ denote the loss for the $j$-th interrupt and silent instance. $N_I$ and $N_S$ are the total counts of interrupt (I1$+$I2) and silent (S1$+$S2) instances, respectively. We subsample silent instances to enforce $N_I = N_S$, maintaining balanced training without additional loss weighting. Fine-tuning uses LoRA~\cite{hu2022lora} (rank~8, $\alpha$=32) for parameter efficiency.

\section{Experiments}

\subsection{Datasets}

\textbf{ESC-50}~\cite{piczak2015esc}\footnote{\url{https://github.com/karolpiczak/ESC-50}} contains 2{,}000 five-second clips spanning 50 environmental sound classes across five categories: animals, natural soundscapes, human non-speech, interior/domestic, and exterior/urban sounds. These categories directly correspond to sounds relevant to DHH users' daily awareness needs~\cite{matthews2006dhh, glasser2017dhh}. We use the standard 5-fold cross-validation protocol for reactive classification and construct proactive evaluation sets with balanced interrupt/silent samples across all four decision types.

\noindent \textbf{Epic-Sounds}~\cite{huh2023epic}\footnote{\url{https://epic-kitchens.github.io/epic-sounds/}} provides 44 audio event classes from egocentric kitchen recordings in the EPIC-KITCHENS dataset. Unlike ESC-50's clean, isolated clips, Epic-Sounds features naturally occurring sounds with concurrent background noise from cooking activities, and ground truth labels include both the target event and the primary foreground activity - making it a challenging test for robustness. We use Epic-Sounds exclusively for evaluation (no training), testing zero-shot transfer of ISM.

\subsection{Training details}

We fine-tune Qwen2-Audio-7B using LoRA~\cite{hu2022lora} (rank~8, $\alpha$=32, dropout~0.1) with AdamW ($\beta_1$=0.9, $\beta_2$=0.95, $\epsilon$=1e-8), learning rate 5e-4 with cosine schedule and 5\% warmup, weight decay 0.1, for 10 epochs. Effective batch size is 384 (batch~24 $\times$ 16-step gradient accumulation). Maximum sequence length is 2{,}048 tokens. Training is performed on 6 NVIDIA H100 GPUs. A fixed random seed ensures reproducibility.

\subsection{Evaluation metrics}

We define proactive-specific metrics to evaluate the model's decision-making:
\emph{interrupt precision} ($P_I$) measures correctness of interrupt decisions;
\emph{silent precision} ($P_S$) measures correctness of silent decisions;
\emph{recall} is measured per decision type - $R_{I1}$ (onset interrupt), $R_{I2}$ (sustained-relevance interrupt), $R_{S1}$ (irrelevance silence), and $R_{S2}$ (de-duplication silence). These metrics capture distinct capabilities: $R_{I1}$ and $R_{I2}$ measure sensitivity to relevant events, $R_{S1}$ measures specificity to irrelevant events, and $R_{S2}$ measures history-aware de-duplication. We additionally report \emph{interrupt F1}, computed as $F_I = 2 \cdot P_I \cdot R_{I1} / (P_I + R_{I1})$, using onset recall ($R_{I1}$) as the primary interrupt recall since onset detection is the operationally critical capability in streaming - detecting the first appearance of a relevant event. $R_{I2}$ is reported separately to assess sustained-relevance performance.

All proactive metrics on ESC-50 are reported as averages over the standard 5-fold cross-validation splits. Epic-Sounds metrics are computed on a single evaluation set (no training is performed on this dataset).

\subsection{Proactive evaluation results}

Table~\ref{tab:proactive} presents proactive evaluation on both ESC-50 and Epic-Sounds. We compare three configurations: \emph{Zero-Shot} (base Qwen2-Audio-7B with proactive prompting), \emph{Reactive SFT} (fine-tuned for classification only, without ISM), and \emph{PALLM} (fine-tuned with ISM).

\noindent \textbf{ESC-50.}
The zero-shot baseline exhibits strong I1 recall (100.0\%) but poor S1 recall (59.9\%) and negligible S2 recall (0.2\%), revealing a fundamental affirmative bias - it interrupts regardless of relevance or history, rendering it unusable for real-world monitoring. Reactive SFT improves silent recall but fails at de-duplication (S2: 10.1\%), the very capability that prevents notification fatigue. PALLM with ISM resolves both failure modes, achieving 99.4\% interrupt precision, 100.0\% S2 recall, and 99.6\% interrupt F1. The perfect de-duplication recall is particularly significant: it demonstrates that ISM learns history-aware suppression as an emergent behavior from the four-state training formulation, a capability absent from all baselines.

For reactive classification, PALLM achieves 94.7\% accuracy on ESC-50 (5-fold cross-validation), matching PANN (94.7\%)~\cite{kong2020panns} and approaching AST (95.7\%)~\cite{gong2021ast} - models purpose-built for classification alone. PALLM achieves this while \emph{simultaneously} supporting proactive capabilities that these specialized models lack entirely.

\noindent \textbf{Epic-Sounds (zero-shot transfer).}
Without any training on Epic-Sounds, PALLM achieves the highest interrupt F1 (67.5), outperforming both Reactive SFT (65.2) and Zero-Shot (66.9). Although the F1 margin over Zero-Shot is narrow, the underlying behavior differs substantially: PALLM balances precision and recall (51.9\% $P_I$, 96.7\% $R_{I1}$), whereas Zero-Shot achieves comparable F1 through near-total over-triggering (50.8\% $P_I$, 98.3\% $R_{I1}$, 3.9\% $R_{S1}$) - behavior that would cause severe notification fatigue in a real proactive system. Reactive SFT takes the opposite extreme: high precision (92.9\% $P_I$) but only 50.2\% $R_{I1}$, missing half of all relevant onsets. This illustrates a key insight about proactivity: classification accuracy alone does not produce usable proactive behavior. Reactive SFT's high S1 recall (96.1\%) reflects domain-mismatch conservatism - it fails to recognize out-of-domain sounds and defaults to silence - not genuine proactive competence. PALLM is the only method that maintains strong onset detection (96.7\% $R_{I1}$) without collapsing into either over-triggering or over-suppression.

These results reveal a principled decomposition of proactive transfer: ISM's temporal-structure decisions (I1 onset detection, 96.7\%) transfer robustly across domains, demonstrating that the four-state formulation captures domain-invariant decision structure. Decisions that depend on acoustic clarity are more domain-sensitive: S1 recall is 10.1\% and I2 recall drops to 41.2\%, as the model - trained exclusively on clean, isolated ESC-50 clips - encounters overlapping kitchen sounds for the first time. Critically, this is not an ISM limitation but a training data one: the zero-shot baseline fares even worse on S1 (3.9\%), and Reactive SFT's high S1 (96.1\%) comes only by sacrificing half its interrupt recall. Domain-diverse training data or calibrated decision thresholds should close this gap while preserving ISM's strong interrupt transfer (Section~\ref{sec:conclusion}). De-duplication (S2) is not evaluated on Epic-Sounds, as the protocol does not carry conversation history across samples.

For reactive classification on all 44 Epic-Sounds classes, PALLM achieves 34.5\% top-1 accuracy (state-of-the-art audio-only: 53.8\%~\cite{huh2023epic}). This gap is expected for an out-of-domain model, but it makes interrupt F1 results all the more notable: ISM's proactive decision structure transfers where fine-grained classification does not, confirming that the four-state formulation captures a level of abstraction above per-class discrimination.

\begin{table}[t]
  \centering
  \resizebox{\linewidth}{!}{
  \begin{tabular}{@{}c l ccc cccc@{}}
    \toprule
    \multirow{2}{*}{\textbf{Data}} & \multirow{2}{*}{\textbf{Method}}
    & \multicolumn{2}{c}{\textbf{Precision}} & \textbf{F1}
    & \multicolumn{4}{c}{\textbf{Recall}} \\
    \cmidrule(lr){3-4} \cmidrule(lr){5-5} \cmidrule(lr){6-9}
    & & $P_I$ & $P_S$ & $F_I$ & $R_{I1}$ & $R_{I2}$ & $R_{S1}$ & $R_{S2}$ \\
    \midrule
    \multirow{3}{*}{\makecell{ESC-50}}
    & Zero-Shot & 58.8 & 99.6 & 74.1 & \textbf{100.0} & 99.8 & 59.9 & 0.2 \\
    & Reactive SFT & 67.2 & 90.1 & 79.8 & 98.1 & 89.9 & 98.1 & 10.1 \\
    & \textbf{PALLM (ISM)} & \textbf{99.4} & \textbf{99.7} & \textbf{99.6} & 99.8 & \textbf{100.0} & \textbf{99.4} & \textbf{100.0} \\
    \midrule
    \multirow{3}{*}{\makecell{Epic-\\Sounds}}
    & Zero-Shot & 50.8 & 69.7 & 66.9 & \textbf{98.3} & \textbf{100.0} & 3.9 & --- \\
    & Reactive SFT & \textbf{92.9} & 65.7 & 65.2 & 50.2 & 76.5 & \textbf{96.1} & --- \\
    & \textbf{PALLM (ISM)} & 51.9 & \textbf{71.9} & \textbf{67.5} & 96.7 & 41.2 & 10.1 & --- \\
    \bottomrule
  \end{tabular}
  }
  
\caption{Proactive evaluation results (\%). $P$: precision, $R$: recall, $F_I$: interrupt F1. I: interrupt, S: silent. Subscripts denote type. On Epic-Sounds, Reactive SFT's high $R_{S1}$ reflects domain-mismatch conservatism (halved $R_{I1}$), not proactive competence. PALLM is the only method maintaining high $R_{I1}$ while achieving the best F1. S2 is excluded on Epic-Sounds (no conversation history).}
\label{tab:proactive}
\vspace{-6mm}

\end{table}

\subsection{Streaming evaluation}

To evaluate interrupt latency and de-duplication under controlled conditions, we construct a streaming protocol using 400 samples spanning all 50 ESC-50 classes (8 per class). Each 15-second composite sample is constructed by concatenating a randomly selected irrelevant audio segment (seconds 0-5), the target relevant segment (seconds 5-10), and another irrelevant segment (seconds 10-15), with faded transitions between segments to avoid temporal artifacts. During streaming, a 5-second rolling audio window is maintained, and the full conversation history is preserved and updated upon each interrupt event. The model makes token-by-token causal predictions - it cannot observe future audio.

\begin{figure}[t]
  \centering
  \includegraphics[width=\linewidth]{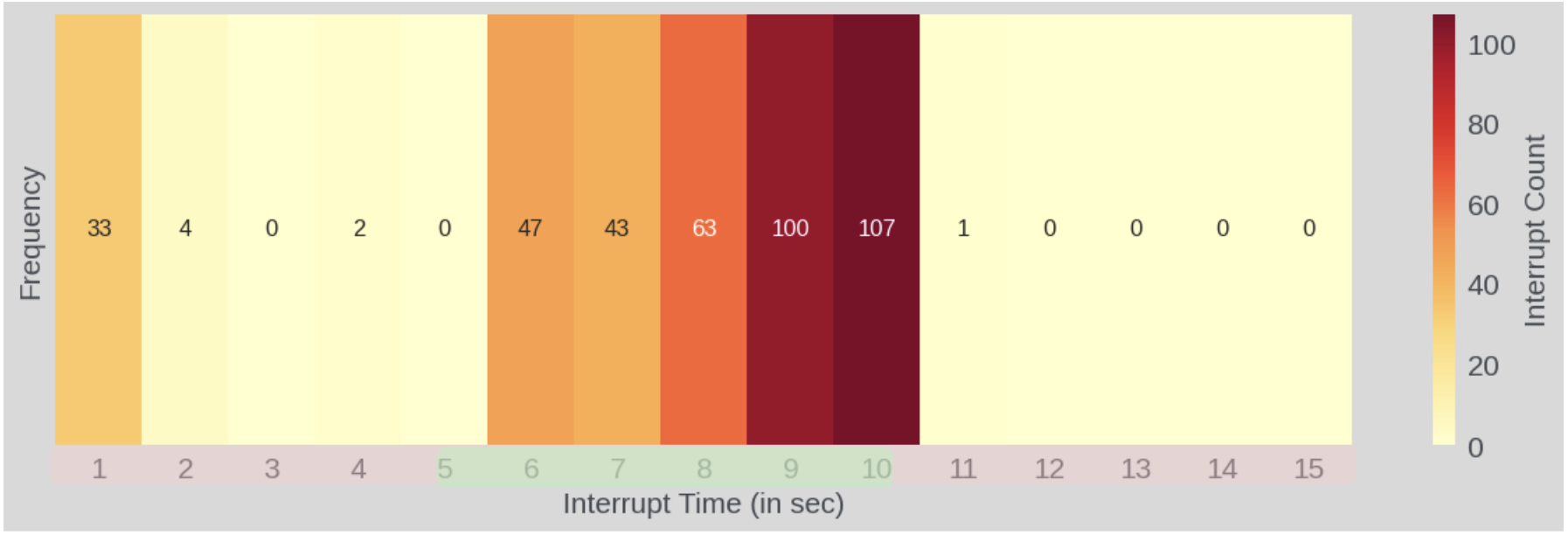}
    \caption{Streaming evaluation on 400 samples across 50 sound classes. Each column represents a 1-second interval; Green region (5-10s) marks when the relevant audio is present. PALLM triggers interrupts within 3.5s of event onset on average.}
  \label{fig:streaming}
\end{figure}

PALLM achieves an average response latency of \textbf{3.5~seconds} from onset of the relevant audio event, with nearly all interrupts occurring within the relevant time window (Figure~\ref{fig:streaming}). This protocol validates two essential properties for deployment: (1)~causal inference produces timely interrupts under real-time constraints, and (2)~conversation-history-based de-duplication suppresses redundant alerts across successive windows. Both are confirmed, establishing that ISM-based proactive monitoring is viable under real-time constraints suitable for wearable DHH applications ~\cite{matthews2006dhh, glasser2017dhh}. The fixed onset position provides a controlled setting for these measurements; future work will randomize onset time and duration and explore the latency-hallucination tradeoff from shorter training segments.

%
%

\subsection{Ablation: effect of Interrupt Type-2}
\label{sec:ablation}

We ablate the contribution of I2 training by comparing models trained with I1-only versus I1+I2. Adding I2 improves I2 recall by ${\sim}$10 percentage points on ESC-50. I1 alone only trains the model to interrupt when relevant audio follows irrelevant audio; if the model misses the initial onset and observes a window of entirely relevant audio, it has no training signal for this case. I2 explicitly addresses this gap, ensuring robust triggering across both onset and sustained-relevance scenarios.

\section{Conclusion and future work}
\label{sec:conclusion}

We introduced proactive audio assistance, a new task for AudioLLMs in which the model monitors an audio stream and autonomously decides when to alert the user from a single watch-out intent. ISM embeds this capability into standard LLM decoding through two special tokens, requiring no architectural changes. On ESC-50, ISM achieves 99.6\% interrupt F1 with perfect de-duplication recall, resolving the notification fatigue problem in existing sound awareness systems. On Epic-Sounds - without any domain-specific training - ISM is the only method that maintains strong onset detection (96.7\% $R_{I1}$) without collapsing into over-triggering or over-suppression, achieving the highest interrupt F1; the remaining silent-recall gap is attributable to training data, not the framework, and is directly addressable through domain-diverse training. Streaming evaluation confirms real-time viability with 3.5s average latency.

Future work includes domain-diverse training to close the silent-recall gap, reducing latency while managing the hallucination tradeoff, extending to implicit and semantic user intents, and user studies with DHH participants~\cite{ohshiro2022dhh}. Since ISM requires only vocabulary extension and autoregressive decoding, it can be applied to any current or future AudioLLM as-is, making proactive audio assistance a portable capability rather than a model-specific feature.

\section{Generative AI use disclosure}

Generative AI tools were used for editing and polishing the manuscript text. All experimental design, implementation, evaluation, and scientific claims are the sole responsibility of the authors.

\bibliographystyle{IEEEtran}
\bibliography{pallm}

@inproceedings{piczak2015esc,
  title={{ESC}: Dataset for Environmental Sound Classification},
  author={Piczak, Karol J.},
  booktitle={Proc. ACM Multimedia},
  pages={1015--1018},
  year={2015}
}

@inproceedings{huh2023epic,
  title={{EPIC-SOUNDS}: A Large-Scale Dataset of Actions That Sound},
  author={Huh, Jaesung and Chalk, Jacob and Kazakos, Evangelos and Damen, Dima and Zisserman, Andrew},
  booktitle={Proc. ICASSP},
  pages={1--5},
  year={2023}
}

@inproceedings{gemmeke2017audioset,
  title={Audio Set: An ontology and human-labeled dataset for audio events},
  author={Gemmeke, Jort F. and Ellis, Daniel P. W. and Freedman, Dylan and Jansen, Aren and Lawrence, Wade and Moore, R. Channing and Plakal, Manoj and Ritter, Marvin},
  booktitle={Proc. ICASSP},
  pages={776--780},
  year={2017}
}

@article{chu2024qwen2audio,
  title={{Qwen2-Audio}: Technical Report},
  author={Chu, Yunfei and Xu, Jin and Yang, Qian and Wei, Haojie and Wei, Xipin and Guo, Zhifang and Leng, Yichong and Lv, Yuanjun and He, Jinzheng and Lin, Junyang and Zhou, Chang and Zhou, Jingren},
  journal={arXiv preprint arXiv:2407.10759},
  year={2024}
}

@inproceedings{ghosh2024gama,
  title={{GAMA}: A Large Audio-Language Model with Advanced Audio Understanding and Complex Reasoning Abilities},
  author={Ghosh, Sreyan and Evuru, Chandra Kiran Reddy and Kumar, Sonal and Sakshi, S. and Tyagi, Utkarsh and Ramaneswaran, S. and Sethuraman, Sharath and Manocha, Dinesh},
  booktitle={Proc. ACL},
  year={2024}
}

@article{he2024meralion,
  title={MERaLiON-AudioLLM: Bridging Audio and Language with Large Language Models},
  author={He, Yingxu and Leong, Zhuohan and others},
  journal={arXiv preprint arXiv:2406.07146},
  year={2024}
}

@article{radford2023whisper,
  title={Robust Speech Recognition via Large-Scale Weak Supervision},
  author={Radford, Alec and Kim, Jong Wook and Xu, Tao and Brockman, Greg and McLeavey, Christine and Sutskever, Ilya},
  journal={Proc. ICML},
  pages={28492--28518},
  year={2023}
}

@inproceedings{gong2021ast,
  title={{AST}: Audio Spectrogram Transformer},
  author={Gong, Yuan and Chung, Yu-An and Glass, James},
  booktitle={Proc. Interspeech},
  pages={571--575},
  year={2021}
}

@article{kong2020panns,
  title={{PANNs}: Large-Scale Pretrained Audio Neural Networks for Audio Pattern Recognition},
  author={Kong, Qiuqiang and Cao, Yin and Iqbal, Turab and Wang, Yuxuan and Wang, Wenwu and Plumbley, Mark D.},
  journal={IEEE/ACM Transactions on Audio, Speech, and Language Processing},
  volume={28},
  pages={2880--2894},
  year={2020}
}

@inproceedings{chen2023beats,
  title={{BEATs}: Audio Pre-Training with Acoustic Tokenizers},
  author={Chen, Sanyuan and Wu, Yu and Wang, Chengyi and Liu, Shujie and Tompkins, Daniel and Chen, Zhuo and Wei, Furu},
  booktitle={Proc. ICML},
  pages={5178--5193},
  year={2023}
}

@inproceedings{chen2022htsat,
  title={{HTS-AT}: A Hierarchical Token-Semantic Audio Transformer for Sound Classification and Detection},
  author={Chen, Ke and Du, Xingjian and Zhu, Bilei and Ma, Zejun and Berg-Kirkpatrick, Taylor and Dubnov, Shlomo},
  booktitle={Proc. ICASSP},
  pages={646--650},
  year={2022}
}

@inproceedings{hu2022lora,
  title={{LoRA}: Low-Rank Adaptation of Large Language Models},
  author={Hu, Edward J. and Shen, Yelong and Wallis, Phillip and Allen-Zhu, Zeyuan and Li, Yuanzhi and Wang, Shean and Wang, Lu and Chen, Weizhu},
  booktitle={Proc. ICLR},
  year={2022}
}

@inproceedings{glasser2017dhh,
  title={Sound Awareness for Deaf and Hard of Hearing Users},
  author={Glasser, Abraham and Kushalnagar, Kesavan and Kushalnagar, Raja},
  booktitle={Proc. ASSETS},
  pages={321--322},
  year={2017}
}

@inproceedings{matthews2006dhh,
  title={A Wearable Sound Awareness System for Deaf Users},
  author={Matthews, Tara and Fong, Janette and Mankoff, Jennifer},
  booktitle={Proc. ASSETS},
  pages={170--177},
  year={2006}
}

@inproceedings{chen2024videollmonline,
  title={{VideoLLM-online}: Online Video Large Language Model for Streaming Video},
  author={Chen, Joya and Ge, Zhaoyang and Zhu, Weiqi and Xie, Rui and Ge, Kevin and Lan, Wayne and Duan, Haoqi and others},
  booktitle={Proc. CVPR},
  pages={18407--18417},
  year={2024}
}

@inproceedings{lee2024mirai,
  title={Can Large Language Models be Good Companions? An LLM-Based Eyewear System with Proactive Dialogue},
  author={Lee, Hao-Nan and Chung, Yi-Chieh and others},
  journal={arXiv preprint arXiv:2401.06700},
  year={2024}
}

@inproceedings{deng2023proactive,
  title={Prompting and Evaluating Large Language Models for Proactive Dialogues: Clarification, Target-guided, and Non-collaboration},
  author={Deng, Yang and Lei, Wenqiang and Lam, Wai and Chua, Tat-Seng},
  booktitle={Findings of EMNLP},
  pages={10602--10621},
  year={2023}
}

@inproceedings{ohshiro2022dhh,
    title={How People Who Are Deaf, {D}eaf, and Hard of Hearing Use Technology in Creative Sound Activities},
    author={Ohshiro, Keita and Cartwright, Mark},
    booktitle={Proc. ASSETS},
    articleno={66},
    numpages={4},
    year={2022}
}

@misc{kundu2026planwatchrecoverbenchmark,
      title={Plan, Watch, Recover: A Benchmark and Architectures for Proactive Procedural Assistance}, 
      author={Kaustav Kundu and Ritvik Shrivastava and Maxim Arap and Nanshu Wang and Xianhui Zhu and Quintin Fettes and Gautam Tiwari and Parth Suresh and Théo Moutakanni and Alejandro Castillejo Munoz and Allen Bolourchi and Pascale Fung and Pinar Donmez and Babak Damavandi and Anuj Kumar and Seungwhan Moon},
      year={2026},
      eprint={2606.04970},
      archivePrefix={arXiv},
      primaryClass={cs.CV},
      url={https://arxiv.org/abs/2606.04970}, 
}

\end{document}